\documentclass{IET-Conf-Paper}

\usepackage{amsmath}
\usepackage{amssymb}
\usepackage{array}
\usepackage{tabularx}
\usepackage{booktabs}
\usepackage[ruled,vlined,linesnumbered]{algorithm2e}
\usepackage{url}
\usepackage[colorlinks=true,linkcolor=blue,citecolor=blue,urlcolor=blue]{hyperref}

\newcommand{\autonumlabel}[1]{%
  \refstepcounter{equation}%
  \label{#1}%
  (\theequation)%
}

\begin{document}

\title{Seeking Cost--Effective Incentive for Merchant MTDC Investment in an AC Grid: A Parametric MISOCP Model to VSC Siting and Sizing}

\author{Haixiao Li\ad{1}\corr, Jure Lokar\ad{2}, Leopold Herman\ad{2}, Aleksandra Lekić\ad{1}}

\address{\add{1}{Department of Electrical Sustainable Energy, Delft University of Technology, Delft, The Netherlands}
\add{2}{Faculty of Electrical Engineering, University of Ljubljana, Ljubljana, Slovenia}
\email{haixiaoli.ee@gmail.com}}


\begin{abstract}
Deploying voltage source converter (VSC)-based multi-terminal DC (MTDC) networks in existing AC grids offers a promising solution for alleviating transmission congestion, enhancing power-transfer capability, and improving operational economy. However, constructing an MTDC network requires participation from merchant investors. Therefore, the transmission system operator (TSO) responsible for the AC grid must provide an adequate economic incentive that enables merchant investors not only to recover the MTDC investment cost but also to earn a reasonable return commensurate with their risk and expectations. An insufficient incentive may fail to attract merchant investment, whereas an excessive incentive may unnecessarily reduce the economic benefits retained by the TSO. Determining an appropriate incentive that balances these competing interests therefore becomes a critical issue. To address this issue, this paper formulates a sharing-rate-parameterized mixed-integer second-order cone programming (MISOCP) model that jointly optimizes VSC siting and sizing while coordinating AC/DC system operation. A straightforward search procedure is then employed to identify the benefit-sharing rate that meets the investor’s expectations while preserving sufficient benefits for the TSO. Case studies demonstrate the operating-cost benefits of MTDC integration and validate the proposed search strategy for seeking such a benefit-sharing rate.
\end{abstract}

\maketitle

\setcounter{footnote}{0}
\renewcommand{\thefootnote}{\arabic{footnote}}

\section{Introduction}
\subsection{Background and Motivation}
Driven by urban redevelopment and growing electricity demand, existing AC grids face an increasing need for transmission expansion. Integrating voltage-source-converter (VSC)-based multi-terminal DC (MTDC) networks offers a promising option to relieve transmission congestion and enhance system loadability without extensive reinforcement of the existing AC network. Through flexible and independent control of active and reactive power at multiple terminals, VSC-MTDC networks can redispatch power flows away from congested corridors \cite{b1}. 

However, these benefits depend strongly on the siting and sizing of the VSCs. Inappropriate connection points may provide limited congestion relief, whereas undersized converters restrict power-transfer capability, and oversized converters result in underutilized investment. Jointly optimizing VSC siting and sizing is therefore essential to realizing the technical and economic value of an MTDC project \cite{b2}.

In practice, however, the transmission system operator (TSO)'s interest alone is insufficient to bring an MTDC project into operation. A merchant investor must commit substantial capital before operation, while much of the resulting benefit accrues to the power system through reduced congestion, lower redispatch costs, and greater operational flexibility. Unless a portion of these system-wide benefits can be allocated to the investor, enabling the recovery of investment costs and the attainment of its expected return while preserving benefits for the TSO, an MTDC project may not be financially viable and, consequently, may not proceed to implementation.

Accordingly, a key issue is to identify a cost-effective incentive that enables merchant investors to make economically viable VSC siting and sizing decisions while ensuring that the TSO responsible for the AC grid retains sufficient economic benefits.

\subsection{Literature Review}
Optimal VSC siting and sizing are critical components of AC/DC system expansion planning and have attracted considerable research attention. Existing studies have commonly addressed VSC siting and sizing through AC/DC optimal power flow (OPF) formulations, in which converter locations and capacities are jointly optimized while satisfying the operational constraints of both AC and DC networks \cite{b3, b4}. Researchers have also devoted considerable attention to the inherent modeling challenges of AC/DC OPF, including convex relaxations of nonlinear AC/DC power-flow equations \cite{b5}, multi-period decision-making \cite{b6}, and the integration of uncertainties in renewable generation \cite{b7}.

Nevertheless, a less explored issue is whether the VSC siting and sizing decisions derived from an OPF-based planning model can actually be implemented. In practice, MTDC facilities may be financed by merchant investors whose objective is to guarantee investment returns rather than system-wide operating benefits. Consequently, a VSC siting and sizing plan that is promising for AC/DC system operation may not provide sufficient financial incentives for the investor.

It is therefore necessary to incorporate the merchant investor’s perspective into the AC/DC OPF-based planning model, so that the resulting VSC siting and sizing decisions are not only technically beneficial to the AC/DC system but also economically attractive for the investor to implement. A direct and effective way to address this misalignment is to share part of the TSO's economic benefits with the merchant investor as an investment incentive. By linking the investor’s revenue to the benefits generated by the MTDC construction project, such a benefit-sharing arrangement can enhance investment attractiveness while aligning the interests of the investor and the TSO \cite{b8}.

This, in turn, shifts the decision perspective back to the TSO. Once a cost-effective incentive is adopted, a meaningful question is how large the benefit-sharing rate should be. A rate that is too low may fail to induce the investor to implement an economically viable VSC siting and sizing plan, whereas an unnecessarily high rate may reduce the economic benefits retained by the TSO. It is therefore important to determine the optimal benefit-sharing rate that satisfies the investor’s expected return requirement while preserving sufficient benefits for the TSO.

\subsection{Contributions and Outlines}
This paper develops an optimization model to determine a benefit-sharing rate that attracts merchant investment in the MTDC network while jointly optimizing VSC siting and sizing. In brief, the proposed model links the economic interests of the TSO and the merchant investor through a benefit-sharing arrangement. This linkage directly affects VSC siting and sizing, since the selected connection points and converter capacities determine both the operating-cost savings achieved by the TSO and the investment cost borne by the investor. The resulting planning problem therefore seeks a balance between maximizing system operating-cost savings and maintaining the economic viability of the MTDC investment. Based on such relationship, the optimal benefit-sharing rate is further determined.

To formulate the VSC siting and sizing problem, binary variables are introduced to represent the candidate AC connection points. The nonlinear AC/DC power-flow equations are then convexified, resulting in a mixed-integer second-order cone programming (MISOCP) model. The benefit-sharing sharing rate is subsequently treated as a key parameter affecting the investor's planning decisions, and the model is extended to a parametric MISOCP formulation. Finally, a straightforward search procedure is developed to solve such a parametric MISOCP problem and seek an optimal benefit-sharing rate. Its effectiveness is demonstrated using a customized AC/DC hybrid system.

The remainder of this paper is organized as follows. Section~\ref{sec:2} formulates the AC/DC power flow constraints and the MTDC investment model, including VSC siting, sizing, and cost-effective incentives. Section~\ref{sec:3}  presents the solution procedure for the resulting parametric MISOCP. Section~\ref{sec:4} reports the case study results. Section~\ref{sec:5} concludes the paper.

\section{Model Formulation}
\label{sec:2}
This section presents the operating constraints associated with the AC/DC hybrid system. For ease of presentation, the main notation used is summarized as follows. $\mathcal{N}/\mathcal{E}$ denote the system-wide node/branch sets. $\mathcal{V}$ denotes the set of VSC stations. $(\bar{\bullet})/(\underline{\bullet})$ denote the upper/lower bounds of optimization variables. $(\bullet)^{\mathrm{ac}}/(\bullet)^{\mathrm{dc}}/(\bullet)^{\mathrm{vsc},v}$ denote the parameters or variables associated with the AC grid, MTDC grid, and the VSC station\#$v$, respectively.
\subsection{AC/DC Power Flow Constraints}
Since VSC siting decisions are binary, we present a convex relaxation of the AC/DC power-flow constraints to improve computational tractability. To facilitate the subsequent derivation and interpretation of the relaxation, we first formulate the AC/DC power-flow constraints for a given MTDC network topology.

\setcounter{equation}{0}

\begin{table*}[!t]
\centering
\caption{Constraints of the AC Grid}
\label{tab:ac_socp}
\renewcommand{\arraystretch}{1.15}
\setlength{\tabcolsep}{4pt}

\begin{tabularx}{\textwidth}{
@{}>{\raggedright\arraybackslash}X
>{\centering\arraybackslash}p{1.2cm}@{}
}
\toprule

\multicolumn{2}{@{}l@{}}{\textit{AC branches}} \\

$\displaystyle
p_{ij}^{\mathrm{ac}}
=
g_{ij}^{\mathrm{ac}}
\left(c_{ii}^{\mathrm{ac}}-c_{ij}^{\mathrm{ac}}\right)
+
b_{ij}^{\mathrm{ac}}s_{ij}^{\mathrm{ac}},
\quad
q_{ij}^{\mathrm{ac}}
=
-b_{ij}^{\mathrm{ac}}
\left(c_{ii}^{\mathrm{ac}}-c_{ij}^{\mathrm{ac}}\right)
+
g_{ij}^{\mathrm{ac}}s_{ij}^{\mathrm{ac}},
\quad
\forall(i,j)\in\mathcal{E}^{\mathrm{ac}}.
$
& \autonumlabel{eq:1} \\

$\displaystyle
-\bar{s}_{ij}^{\mathrm{ac}}
\leq
\cos\left(\frac{k\pi}{N}\right)p_{ij}^{\mathrm{ac}}
+
\sin\left(\frac{k\pi}{N}\right)q_{ij}^{\mathrm{ac}}
\leq
\bar{s}_{ij}^{\mathrm{ac}},
\quad
\forall k\in\{1,\ldots,N\},
\quad
\forall(i,j)\in\mathcal{E}^{\mathrm{ac}}.
$
& \autonumlabel{eq:2} \\

\addlinespace[0.2em]
\multicolumn{2}{@{}l@{}}{\textit{AC buses}} \\

$\displaystyle
p_i^{\mathrm{ac}}
=
\sum_{j:(i,j)\in\mathcal{E}^{\mathrm{ac}}}
p_{ij}^{\mathrm{ac}}
+
G_{i,\mathrm{sh}}^{\mathrm{ac}}c_{ii}^{\mathrm{ac}},
\quad
q_i^{\mathrm{ac}}
=
\sum_{j:(i,j)\in\mathcal{E}^{\mathrm{ac}}}
q_{ij}^{\mathrm{ac}}
-
B_{i,\mathrm{sh}}^{\mathrm{ac}}c_{ii}^{\mathrm{ac}},
\quad
\forall i\in\mathcal{N}^{\mathrm{ac}}.
$
& \autonumlabel{eq:3} \\

$\displaystyle
p_i^{\mathrm{ac}}
=
p_{i,\mathrm{gen}}^{\mathrm{ac}}
-
p_{i,\mathrm{load}}^{\mathrm{ac}}
-
p_{i,\mathrm{a2v}}^{\mathrm{ac}},
\quad
q_i^{\mathrm{ac}}
=
q_{i,\mathrm{gen}}^{\mathrm{ac}}
-
q_{i,\mathrm{load}}^{\mathrm{ac}}
-
q_{i,\mathrm{a2v}}^{\mathrm{ac}},
\quad
\forall i\in\mathcal{N}^{\mathrm{ac}}.
$
& \autonumlabel{eq:4} \\

$\displaystyle
\left(\underline{v}_i^{\mathrm{ac}}\right)^2
\leq
c_{ii}^{\mathrm{ac}}
\leq
\left(\bar{v}_i^{\mathrm{ac}}\right)^2,
\quad
\forall i\in\mathcal{N}^{\mathrm{ac}}.
$
& \autonumlabel{eq:5} \\

\addlinespace[0.2em]
\multicolumn{2}{@{}l@{}}{\textit{Generators}} \\

$\displaystyle
\underline{p}_{i,\mathrm{gen}}^{\mathrm{ac}}
\leq
p_{i,\mathrm{gen}}^{\mathrm{ac}}
\leq
\bar{p}_{i,\mathrm{gen}}^{\mathrm{ac}},
\quad
\underline{q}_{i,\mathrm{gen}}^{\mathrm{ac}}
\leq
q_{i,\mathrm{gen}}^{\mathrm{ac}}
\leq
\bar{q}_{i,\mathrm{gen}}^{\mathrm{ac}},
\quad
\forall i\in\mathcal{N}^{\mathrm{ac}}.
$
& \autonumlabel{eq:6} \\

\addlinespace[0.2em]
\multicolumn{2}{@{}l@{}}{\textit{SOCP relaxation}} \\

$\displaystyle
\left(c_{ij}^{\mathrm{ac}}\right)^2
+
\left(s_{ij}^{\mathrm{ac}}\right)^2
\leq
c_{ii}^{\mathrm{ac}}c_{jj}^{\mathrm{ac}},
\quad
\forall(i,j)\in\mathcal{E}^{\mathrm{ac}}.
$
& \autonumlabel{eq:7} \\

\bottomrule
\end{tabularx}
\end{table*}

\begin{table*}[!t]
\centering
\caption{Constraints of the MTDC Network}
\label{tab:dc_socp}
\renewcommand{\arraystretch}{1.08}
\setlength{\tabcolsep}{3pt}

\begin{tabularx}{\textwidth}{
@{}>{\raggedright\arraybackslash}X
>{\centering\arraybackslash}p{1.2cm}@{}
}
\toprule

\multicolumn{2}{@{}l@{}}{\textit{DC branches}} \\

$\displaystyle
p_{jk}^{\mathrm{dc}}
+
p_{kj}^{\mathrm{dc}}
=
r_{jk}^{\mathrm{dc}}l_{jk}^{\mathrm{dc}},
\quad
\forall(j,k)\in\mathcal{E}^{\mathrm{dc}}.
$
& \autonumlabel{eq:8} \\

$\displaystyle
u_j^{\mathrm{dc}}
-
u_k^{\mathrm{dc}}
=
r_{jk}^{\mathrm{dc}}
\left(p_{jk}^{\mathrm{dc}}-p_{kj}^{\mathrm{dc}}\right),
\quad
\forall(j,k)\in\mathcal{E}^{\mathrm{dc}}.
$
& \autonumlabel{eq:9} \\

$\displaystyle
\underline{p}_{jk}^{\mathrm{dc}}
\leq
p_{jk}^{\mathrm{dc}}
\leq
\bar{p}_{jk}^{\mathrm{dc}},
\quad
\forall(j,k)\in\mathcal{E}^{\mathrm{dc}}.
$
& \autonumlabel{eq:10} \\

\addlinespace[0.2em]
\multicolumn{2}{@{}l@{}}{\textit{DC buses}} \\

$\displaystyle
p_j^{\mathrm{dc}}
=
\rho^{\mathrm{dc}}
\sum_{k:(j,k)\in\mathcal{E}^{\mathrm{dc}}}
p_{jk}^{\mathrm{dc}},
\quad
\forall j\in\mathcal{N}^{\mathrm{dc}}.
$
& \autonumlabel{eq:11} \\

$\displaystyle
\left(\underline{v}_j^{\mathrm{dc}}\right)^2
\leq
u_j^{\mathrm{dc}}
\leq
\left(\bar{v}_j^{\mathrm{dc}}\right)^2,
\quad
\forall j\in\mathcal{N}^{\mathrm{dc}}.
$
& \autonumlabel{eq:12} \\

\addlinespace[0.2em]
\multicolumn{2}{@{}l@{}}{\textit{SOCP relaxation}} \\

$\displaystyle
\left(p_{jk}^{\mathrm{dc}}\right)^2
\leq
l_{jk}^{\mathrm{dc}}u_j^{\mathrm{dc}},
\quad
\forall(j,k)\in\mathcal{E}^{\mathrm{dc}}.
$
& \autonumlabel{eq:13} \\

\bottomrule
\end{tabularx}
\vspace{-4pt}
\end{table*}

\begin{table*}[!t]
\centering
\caption{Constraints of the VSC Station\#$v$}
\label{tab:vsc_socp}
\renewcommand{\arraystretch}{1.08}
\setlength{\tabcolsep}{3pt}

\begin{tabularx}{\textwidth}{
@{}>{\raggedright\arraybackslash}X
>{\centering\arraybackslash}p{1.2cm}@{}
}
\toprule

\multicolumn{2}{@{}l@{}}{\textit{VSC buses}} \\

$\displaystyle
p_s^{\mathrm{vsc},v}
=
g_{sc}^{\mathrm{vsc},v}
\left(c_{ss}^{\mathrm{vsc},v}-c_{sc}^{\mathrm{vsc},v}\right)
+
b_{sc}^{\mathrm{vsc},v}s_{sc}^{\mathrm{vsc},v},
\quad
q_s^{\mathrm{vsc},v}
=
-b_{sc}^{\mathrm{vsc},v}
\left(c_{ss}^{\mathrm{vsc},v}-c_{sc}^{\mathrm{vsc},v}\right)
+
g_{sc}^{\mathrm{vsc},v}s_{sc}^{\mathrm{vsc},v},
\quad
\forall v\in\mathcal{V},
\quad
(s,c)\in\mathcal{E}^{\mathrm{vsc},v}.
$
& \autonumlabel{eq:14} \\

$\displaystyle
p_c^{\mathrm{vsc},v}
=
g_{cs}^{\mathrm{vsc},v}
\left(c_{cc}^{\mathrm{vsc},v}-c_{cs}^{\mathrm{vsc},v}\right)
+
b_{cs}^{\mathrm{vsc},v}s_{cs}^{\mathrm{vsc},v},
\quad
q_c^{\mathrm{vsc},v}
=
-b_{cs}^{\mathrm{vsc},v}
\left(c_{cc}^{\mathrm{vsc},v}-c_{cs}^{\mathrm{vsc},v}\right)
+
g_{cs}^{\mathrm{vsc},v}s_{cs}^{\mathrm{vsc},v},
\quad
\forall v\in\mathcal{V},
\quad
(c,s)\in\mathcal{E}^{\mathrm{vsc},v}.
$
& \autonumlabel{eq:15} \\

$\displaystyle
\left(\underline{v}_c^{\mathrm{vsc},v}\right)^2
\leq
c_{cc}^{\mathrm{vsc},v}
\leq
\left(\bar{v}_c^{\mathrm{vsc},v}\right)^2,
\quad
\forall v\in\mathcal{V},
\quad
c\in\mathcal{N}^{\mathrm{vsc},v}.
$
& \autonumlabel{eq:16} \\

$\displaystyle
-\bar{s}_c^{\mathrm{vsc},v}
\leq
\cos\left(\frac{k\pi}{N}\right)p_c^{\mathrm{vsc},v}
+
\sin\left(\frac{k\pi}{N}\right)q_c^{\mathrm{vsc},v}
\leq
\bar{s}_c^{\mathrm{vsc},v},
\quad
\forall k\in\{1,\ldots,N\},
\quad
\forall v\in\mathcal{V},
\quad
c\in\mathcal{N}^{\mathrm{vsc},v}.
$
& \autonumlabel{eq:17} \\

\addlinespace[0.2em]
\multicolumn{2}{@{}l@{}}{\textit{VSC power loss}} \\

$\displaystyle
p_c^{\mathrm{vsc},v}
+
p_{\mathrm{loss},c}^{\mathrm{vsc},v}
+
p_j^{\mathrm{dc}}
=
0,
\quad
\forall(v,j)\in\mathcal{A}^\mathrm{v2d},
\quad
c\in\mathcal{N}^{\mathrm{vsc},v}.
$
& \autonumlabel{eq:18} \\

$\displaystyle
p_{\mathrm{loss},c}^{\mathrm{vsc},v}
=
a_{\mathrm{loss}}^{\mathrm{vsc},v}
+
b_{\mathrm{loss}}^{\mathrm{vsc},v}I_c^{\mathrm{vsc},v}
+
c_{\mathrm{loss}}^{\mathrm{vsc},v}l_c^{\mathrm{vsc},v},
\quad
\forall v\in\mathcal{V},
\quad
c\in\mathcal{N}^{\mathrm{vsc},v}.
$
& \autonumlabel{eq:19} \\

$\displaystyle
0
\leq
I_c^{\mathrm{vsc},v}
\leq
\bar{I}_c^{\mathrm{vsc},v},
\qquad
0
\leq
l_c^{\mathrm{vsc},v}
\leq
\left(\bar{I}_c^{\mathrm{vsc},v}\right)^2,
\quad
\forall v\in\mathcal{V},
\quad
c\in\mathcal{N}^{\mathrm{vsc},v}.
$
& \autonumlabel{eq:20} \\

\addlinespace[0.2em]
\multicolumn{2}{@{}l@{}}{\textit{VSC AC terminal}} \\

$\displaystyle
c_{ss}^{\mathrm{vsc},v}
=
c_{ii}^{\mathrm{ac}},
\qquad
p_s^{\mathrm{vsc},v}
=
p_{i,\mathrm{a2v}}^{\mathrm{ac}},
\qquad
q_s^{\mathrm{vsc},v}
=
q_{i,\mathrm{a2v}}^{\mathrm{ac}},
\quad
\forall v\in\mathcal{V},
\quad
\forall(v,i)\in\mathcal{A}^{\mathrm{a2v}},
\quad
s\in\mathcal{N}^{\mathrm{vsc},v}.
$
& \autonumlabel{eq:21} \\

\addlinespace[0.2em]
\multicolumn{2}{@{}l@{}}{\textit{SOCP relaxation}} \\

$\displaystyle
\left(c_{sc}^{\mathrm{vsc},v}\right)^2
+
\left(s_{sc}^{\mathrm{vsc},v}\right)^2
\leq
c_{ss}^{\mathrm{vsc},v}c_{cc}^{\mathrm{vsc},v},
\qquad
\left(I_c^{\mathrm{vsc},v}\right)^2
\leq
l_c^{\mathrm{vsc},v},
\qquad
\left(p_c^{\mathrm{vsc},v}\right)^2
+
\left(q_c^{\mathrm{vsc},v}\right)^2
\leq
c_{cc}^{\mathrm{vsc},v}l_c^{\mathrm{vsc},v},
\quad
\forall v\in\mathcal{V},
\quad
(s,c)\in\mathcal{E}^{\mathrm{vsc},v}.
$
& \autonumlabel{eq:22} \\

\bottomrule
\end{tabularx}
\vspace{-4pt}
\end{table*}

\textit{AC Grid}: The operating constraints of the AC grid are summarized in Table~\ref{tab:ac_socp}. The active and reactive power flows of each AC branch are expressed by Eq.~\eqref{eq:1}, where $g_{ij}^{\mathrm{ac}}$ and $b_{ij}^{\mathrm{ac}}$ denote the branch conductance and susceptance, respectively. The branch apparent-power limit is represented in Eq.~\eqref{eq:2} using an $N$-side polygonal approximation \cite{b10}. The active and reactive power injections at each AC bus are calculated by Eq.~\eqref{eq:3}, including the shunt conductance $G_{i,\mathrm{sh}}^{\mathrm{ac}}$ and shunt susceptance $B_{i,\mathrm{sh}}^{\mathrm{ac}}$. The nodal power balance is imposed by Eq.~\eqref{eq:4}, where $p_{i,\mathrm{a2v}}^{\mathrm{ac}}$ and $q_{i,\mathrm{a2v}}^{\mathrm{ac}}$ represent the powers transferred from AC bus $i$ to a connected VSC station. At buses without generators, $p_{i,\mathrm{gen}}^{\mathrm{ac}}=q_{i,\mathrm{gen}}^{\mathrm{ac}}=0$. At buses without a connected VSC, $p_{i,\mathrm{a2v}}^{\mathrm{ac}}=q_{i,\mathrm{a2v}}^{\mathrm{ac}}=0$. The bus-voltage and generator-output limits are imposed by Eqs.~\eqref{eq:5} and \eqref{eq:6}, respectively. Finally, the original equality that $(c_{ij}^{\mathrm{ac}})^2+(s_{ij}^{\mathrm{ac}})^2=c_{ii}^{\mathrm{ac}}c_{jj}^{\mathrm{ac}}$,
is relaxed to relaxed into the SOCP constraint in Eq.~\eqref{eq:7} \cite{b11}.

\textit{MTDC Network}: The operating constraints of the MTDC network are presented in Table~\ref{tab:dc_socp}. Variables $p_{jk}^{\mathrm{dc}}$ and $p_{kj}^{\mathrm{dc}}$ represent the powers leaving the two ends of DC branch $(j,k)$. Their sum equals the resistive branch loss $r_{jk}^{\mathrm{dc}}l_{jk}^{\mathrm{dc}}$, as described by Eq.~\eqref{eq:8}. The squared-voltage drop and power-transfer limits of the DC branch are imposed by Eqs.~\eqref{eq:9} and \eqref{eq:10}, respectively. The DC nodal power balance is given by Eq.~\eqref{eq:11}, where $\rho^{\mathrm{dc}}$ is the scaling coefficient associated with the adopted DC pole configuration. The DC-bus voltage limits are imposed by Eq.~\eqref{eq:12}. Finally, Eq.~\eqref{eq:13} provides the SOCP relaxation of the original equality that $\left(p_{jk}^{\mathrm{dc}}\right)^2=l_{jk}^{\mathrm{dc}}u_j^{\mathrm{dc}}$ \cite{b12}.

\begin{figure}[h!]
    \centering
    \includegraphics[width=1\linewidth]{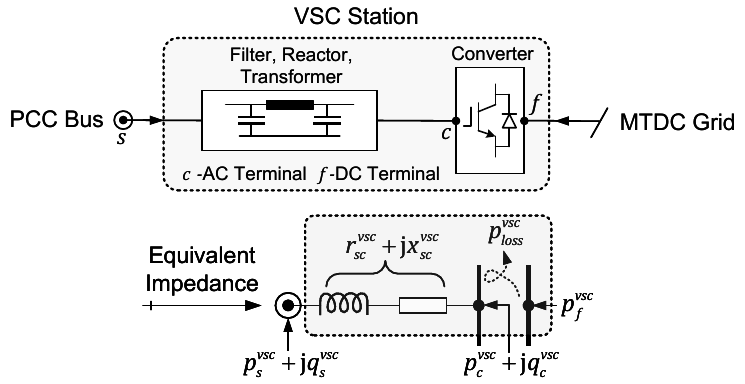}
    \caption{Equivalent impedance model of the VSC station, where the AC-side filter is omitted owing to the low harmonic content of modern modular multilevel structure.}
    \label{fig:1}
    \vspace{2mm}
\end{figure}

\textit{VSC Station}: The operating constraints of the VSC stations are summarized in Table~\ref{tab:vsc_socp}. As presented in Fig.~\ref{fig:1}, each VSC station is represented by a two-bus equivalent circuit, where $s$ and $c$ denote its grid-side and converter-side buses, respectively, and $v\in\mathcal{V}$ identifies the associated VSC station. The active and reactive power flows at the two ends of the equivalent transformer branch of VSC $\#v$ are described by Eqs.~\eqref{eq:14} and \eqref{eq:15}. The converter-side voltage limits are imposed by Eq.~\eqref{eq:16}. Similar to Eq.~\eqref{eq:2}, the circular VSC capability region is represented by an $N$-side polygonal approximation in Eq.~\eqref{eq:17}, which bounds $p_c^{\mathrm{vsc},v}$ and $q_c^{\mathrm{vsc},v}$ by the rated capacity $\bar{s}_c^{\mathrm{vsc},v}$. The active power balance among the converter-side AC power, converter loss, and the power injected into the corresponding DC node $j(v)$ is enforced by Eq.~\eqref{eq:18}. Here, $\mathcal{A}^{\mathrm{v2d}}\subseteq\mathcal{V}\times\mathcal{N}^{\mathrm{dc}}$ denotes the set of fixed VSC--DC node connections. The converter loss is represented by Eq.~\eqref{eq:19} with loss coefficients $a_{\mathrm{loss}}$, $b_{\mathrm{loss}}$, and $c_{\mathrm{loss}}$,  while the converter-current limits are imposed by Eq.~\eqref{eq:20}. For a predetermined AC connection, the coupling between the grid-side bus of VSC $\#v$ and its connected AC bus $i(v)$ is established by Eq.~\eqref{eq:21}. Here, $\mathcal{A}^{\mathrm{a2v}}\subseteq\mathcal{N}^{\mathrm{ac}}\times\mathcal{V}$ denotes the set of fixed AC--VSC node connections. Finally, Eq.~\eqref{eq:22} provides the SOCP relaxations regarding $\left(c_{sc}^{\mathrm{vsc},v}\right)^2
+
\left(s_{sc}^{\mathrm{vsc},v}\right)^2
=
c_{ss}^{\mathrm{vsc},v}c_{cc}^{\mathrm{vsc},v}$, $\left(I_c^{\mathrm{vsc},v}\right)^2
=
l_c^{\mathrm{vsc},v}$, $\left(p_c^{\mathrm{vsc},v}\right)^2
+
\left(q_c^{\mathrm{vsc},v}\right)^2
=
c_{cc}^{\mathrm{vsc},v}l_c^{\mathrm{vsc},v}$ \cite{b13}.

\subsection{VSC Siting and Sizing Constraints}

The merchant investor is assumed to have predetermined the MTDC configuration, including the number of VSCs, the DC topology, the DC-line parameters, and the VSC--DC node connections. This represents, for example, a project with preselected cable corridors and specifications, while AC connection points and converter ratings remain to be determined. The planning, therefore, focuses on VSC siting and sizing.

To model VSC siting, let $z_{vi}$ be a binary variable that $z_{vi}=1$ if VSC $v$ is connected to AC bus $i$ and $z_{vi}=0$ otherwise. Each VSC must be connected to exactly one candidate AC bus:
\begin{equation}
\sum_{i\in\mathcal{N}^{\mathrm{ac}}}z_{vi}=1,
\qquad z_{vi}\in\{0,1\},
\qquad \forall v\in\mathcal{V}.
\label{eq:23}
\end{equation}
To avoid installing an excessive number of converters at the same AC bus, the following optional siting restriction is imposed:
\begin{equation}
\sum_{v\in\mathcal{V}}z_{vi}
\leq 1,
\qquad \forall i\in\mathcal{N}^{\mathrm{ac}}.
\label{eq:24}
\end{equation}
The siting variables must be further coupled with the power exchanged through each candidate AC--VSC connection. Accordingly, the following big-M constraints are introduced to force the exchanged active and reactive powers $p_{vi}^{\mathrm{a2v}}$, $q_{vi}^{\mathrm{a2v}}$ to zero when a candidate connection is not selected:
\begin{equation}
-\mathrm{M}z_{vi}
\leq p_{vi}^{\mathrm{a2v}}
\leq \mathrm{M}z_{vi},
\qquad
\forall v\in\mathcal{V},\
\forall i\in\mathcal{N}^{\mathrm{ac}},
\label{eq:25}
\end{equation}
\begin{equation}
-\mathrm{M}z_{vi}
\leq q_{vi}^{\mathrm{a2v}}
\leq \mathrm{M}z_{vi},
\qquad
\forall v\in\mathcal{V},\
\forall i\in\mathcal{N}^{\mathrm{ac}},
\label{eq:26}
\end{equation}
where $\mathrm{M}>0$ is a sufficiently large positive constant. When $z_{vi}=1$, $p_{vi}^{\mathrm{a2v}}$ and $q_{vi}^{\mathrm{a2v}}$ remain continuous decision variables. Otherwise, both are fixed to zero.

Further, the AC--VSC coupling originally represented by Eqs.~\eqref{eq:4} and \eqref{eq:21} must account for the selected connection points. Specifically, the nodal power balance in Eq.~\eqref{eq:4} retains its original form, while its AC--VSC power exchanges are determined by the siting decisions. The fixed coupling relationships in Eq.~\eqref{eq:21} are replaced by constraints that enforce power and voltage consistencies only for the selected AC--VSC connections. Accordingly, the power-exchange terms in Eq.~\eqref{eq:4} are updated to:
\begin{equation}
p_{i,\mathrm{a2v}}^{\mathrm{ac}}
=
\sum_{v\in\mathcal{V}}p_{vi}^{\mathrm{a2v}},
\qquad
q_{i,\mathrm{a2v}}^{\mathrm{ac}}
=
\sum_{v\in\mathcal{V}}q_{vi}^{\mathrm{a2v}},
\qquad
\forall i\in\mathcal{N}^{\mathrm{ac}}.
\label{eq:27}
\end{equation}
The corresponding grid-side powers of each VSC are given by:
\begin{equation}
p_s^{\mathrm{vsc},v}
=
\sum_{i\in\mathcal{N}^{\mathrm{ac}}}p_{vi}^{\mathrm{a2v}},
\qquad
q_s^{\mathrm{vsc},v}
=
\sum_{i\in\mathcal{N}^{\mathrm{ac}}}q_{vi}^{\mathrm{a2v}},
\qquad
\forall v\in\mathcal{V}.
\label{eq:28}
\end{equation}
Together with the siting and power-activation constraints in Eqs.~\eqref{eq:23}--\eqref{eq:26}, Eqs.~\eqref{eq:27} and \eqref{eq:28} replace the power coupling in Eq.~\eqref{eq:21} for the selected AC--VSC connection. The corresponding voltage coupling is imposed through the following big-M constraint:
\begin{equation}
\begin{aligned}
-\mathrm{M}(1-z_{vi})
&\leq c_{ss}^{\mathrm{vsc},v}-c_{ii}^{\mathrm{ac}}\\
&\leq \mathrm{M}(1-z_{vi}),
\end{aligned}
\quad
\forall v\in\mathcal{V},\
\forall i\in\mathcal{N}^{\mathrm{ac}}.
\label{eq:29}
\end{equation}
When $z_{vi}=1$, Eq.~\eqref{eq:29} enforces $c_{ss}^{\mathrm{vsc},v}=c_{ii}^{\mathrm{ac}}$. Otherwise, the voltage equality is relaxed, and there is no specific relation between $c_{ss}^{\mathrm{vsc},v}$ and $c_{ii}^{\mathrm{ac}}$. Eventually, Eqs.~\eqref{eq:23}--\eqref{eq:29} replace Eq.~\eqref{eq:21} in the planning model.

Regarding VSC sizing, only a minor modification to the operating formulation is required. The capacity $\bar{s}_c^{\mathrm{vsc},v}$ in Eq.~\eqref{eq:17}, previously treated as a constant, is now treated as a decision variable and denoted by $s_c^{\mathrm{vsc},v}$. The associated VSC investment cost, which increases with its rated capacity, is incorporated into the planning objective. This variable represents the minimum converter
capacity required by system operation. Since the rated MVA ratings of VSCs are generally specified as integers in practice, the optimized capacity is rounded up for investment costing. Accordingly, an auxiliary integer variable $n_c^{\mathrm{vsc},v}$ is introduced and constrained by:
\begin{equation}
n_c^{\mathrm{vsc},v}-1+\epsilon
\leq s_c^{\mathrm{vsc},v}
\leq n_c^{\mathrm{vsc},v},
\qquad
\forall v\in\mathcal{V},
\label{eq:30}
\end{equation}
where $\epsilon>0$ is a sufficiently small constant. Thus, $n_c^{\mathrm{vsc},v}=\lceil s_c^{\mathrm{vsc},v}\rceil$ represents the eventually installed capacity of VSC $\#v$ used for costing.

\subsection{Investment Incentive Constraints and Objective Function}
This study considers a simplified incentive mechanism that captures the basic economic conditions for merchant investment. Although practical policies and contractual arrangements may be more complex, the following constraints provide a plausible framework in which the TSO offers incentives and the investor has desire to participate. To evaluate annual economic benefits, representative temporal scenarios are indexed by $t\in\mathcal{T}$. The previously defined operating variables retain their meaning and are indexed by $t$, with operating constraints enforced for each scenario. The VSC siting variables $z_{vi}$ and sizing variables $s_c^{\mathrm{vsc},v}$, $n_c^{\mathrm{vsc},v}$ are shared across all scenarios and therefore carry no scenario index $t$.

\textit{Annual Operating-Cost Savings}:
The annual generation cost with MTDC is evaluated over representative temporal scenarios:
\begin{equation}
C_{\mathrm{op,gen}}^{\mathrm{ann}}
=
\sum_{t\in\mathcal{T}}h_t
\sum_{i\in\mathcal{N}^{\mathrm{ac}}}
\left[
a_i^{\mathrm{ac}}
\left(p_{i,\mathrm{gen},t}^{\mathrm{ac}}\right)^2
+b_i^{\mathrm{ac}}p_{i,\mathrm{gen},t}^{\mathrm{ac}}
+c_i^{\mathrm{ac}}
\right],
\end{equation}
where $h_t$ denotes the hours represented by scenario $t$, with $\sum_{t\in\mathcal{T}}h_t=8760$. The generation-cost coefficients are $a_i^{\mathrm{ac}}$, $b_i^{\mathrm{ac}}$, and $c_i^{\mathrm{ac}}$. The annual savings are defined as:
\begin{equation}
\Delta C_{\mathrm{gen}}^{\mathrm{ann}}
=
C_{0,\mathrm{gen}}^{\mathrm{ann}}
-
C_{\mathrm{op,gen}}^{\mathrm{ann}},
\end{equation}
where $C_{0,\mathrm{gen}}^{\mathrm{ann}}$ is the minimum annual generation cost without MTDC, calculated using the same cost expression and seasonal scenarios, but only contains AC-grid-side constraints Eqs.~\eqref{eq:1}--\eqref{eq:7}, with all AC--VSC power exchanges $p_{i,\mathrm{a2v}}^{\mathrm{ac}}$ and $q_{i,\mathrm{a2v}}^{\mathrm{ac}}$ fixed to zero.

\textit{Investor-Participation Constraint}:
The investor receives a fraction with benefit-sharing $\beta\in[0,1]$ of the annual operating-cost savings. This revenue must cover the annualized investment cost and provide the minimum required annual profit:
\begin{equation}
\begin{aligned}
\beta\Delta C_{\mathrm{gen}}^{\mathrm{ann}}
-
C_{\mathrm{inv}}^{\mathrm{ann}}
&\geq
\underline{\Pi}_{\mathrm{inv}},
\\[-0.1em]
\Longleftrightarrow\quad
\beta C_{\mathrm{op,gen}}^{\mathrm{ann}}
+C_{\mathrm{inv}}^{\mathrm{ann}}
&\leq
\beta C_{0,\mathrm{gen}}^{\mathrm{ann}}
-\underline{\Pi}_{\mathrm{inv}},
\end{aligned}
\label{eq:33}
\end{equation}
where $\underline{\Pi}_{\mathrm{inv}}$ denotes the minimum acceptable annual profit for the investor.

The annual investment cost is:
\begin{equation}
C_{\mathrm{inv}}^{\mathrm{ann}}
=
K^{\mathrm{dc}}
+
\sum_{v\in\mathcal{V}}
\left(
K^{\mathrm{vsc},v}
+
\kappa^{\mathrm{vsc},v}n_c^{\mathrm{vsc},v}
\right),
\label{eq:34}
\end{equation}
where $K^{\mathrm{dc}}$ is the annual fixed cost of the DC network, $K^{\mathrm{vsc},v}$ is the annual fixed cost of VSC $\#v$ with $\kappa^{\mathrm{vsc},v}$ as its annual unit-capacity cost.
We would like to address that Eq.~\eqref{eq:33} is a convex quadratic inequality because $h_t\geq0$ and $a_i^{\mathrm{ac}}\geq0$, while $C_{\mathrm{inv}}^{\mathrm{ann}}$ is affine. Its quadratic terms admit MISOCP representations.

\textit{Objective Function}:
For a given benefit-sharing rate $\beta$, the planning objective is to maximize the annual benefit retained by the TSO:
\begin{equation}
F(\beta)=
\max_{\{z_{vi}\},\,\{s_c^{\mathrm{vsc},v}\},\,
\{n_c^{\mathrm{vsc},v}\},\,
\{\boldsymbol{x}_t\}_{t\in\mathcal{T}}}
\quad
(1-\beta)\Delta C_{\mathrm{gen}}^{\mathrm{ann}}.
\label{eq:inner_planning}
\end{equation}
subject to the operating, siting, sizing, and investor participation constraints, as Eqs.~\eqref{eq:1}--\eqref{eq:20}, \eqref{eq:22}--\eqref{eq:34} introduced above. Here, $\{z_{vi}\}$ denotes the VSC siting variables, $\{s_c^{\mathrm{vsc},v}\}$ denotes the continuous VSC capacity variables, and $\{n_c^{\mathrm{vsc},v}\}$ denotes the integer capacities used for investment costing. These planning variables are shared across all temporal scenarios, whereas $\{\boldsymbol{x}_t\}_{t\in\mathcal{T}}$ comprises the operating variables optimized separately for each scenario $t$.

\subsection{Parametric Optimization Formulation}
The benefit-sharing rate $\beta$ affects both investor participation and the benefit retained by the TSO. For each prescribed $\beta$, Eq.~\eqref{eq:inner_planning}, together with the preceding constraints, defines the inner planning problem. Its optimal value, $F(\beta)$, is the maximum annual TSO benefit achievable by optimizing VSC siting, sizing, and operation at that sharing rate.

The outer parameter-selection problem and the inner planning problem in Eq.~\eqref{eq:inner_planning} are jointly expressed as:
\begin{equation}
\begin{aligned}
&\text{Outer Problem:}\\[-0.2em]
&\quad \beta^\star\in
\operatorname*{arg\,max}_{\beta\in[0,1]}
F(\beta),\\[0.5em]
&\text{Inner Problem (given $\hat{\beta}$):}\\[-0.2em]
&\quad F(\hat{\beta})=
\max_{\{z_{vi}\},\,\{s_c^{\mathrm{vsc},v}\},\,\{n_c^{\mathrm{vsc},v}\},\{\boldsymbol{x}_t\}_{t\in\mathcal{T}}}
(1-\hat{\beta})\Delta C_{\mathrm{gen}}^{\mathrm{ann}}.
\end{aligned}
\label{eq:outer_incentiv}
\end{equation}
The inner problem is subject to  Eqs.~\eqref{eq:1}--\eqref{eq:20}, \eqref{eq:22}--\eqref{eq:34}. If multiple $\beta$ reach the same maximum benefit, the smallest $\beta$ is selected. The resulting optimal $\beta^\star$ and associated VSC configuration related to $\bigl(\{z_{vi}^{\star}\},\,\{s_c^{\mathrm{vsc},v\star}\},\,\{n_c^{\mathrm{vsc},v\star}\}\bigr)$ maximize the TSO's retained benefit while satisfying the investor's minimum profit requirement. Eq.~\eqref{eq:outer_incentiv} therefore defines a parametric MISOCP formulation. Such a formulation cannot be handled directly via off-the-shelf optimizers, and the solution procedure is presented in the next section.

\section{Problem Solving}
\label{sec:3}
To solve the parametric MISOCP in Eq.~\eqref{eq:outer_incentiv}, a straightforward solution strategy is proposed. First, for each fixed benefit-sharing rate $\hat\beta$, the inner problem is solved as a standard MISOCP, and the resulting TSO benefit is recorded. Then, a heuristic search can be used to find the best $\beta^\star$. Specifically, several sharing rates are first tested across $[0,1]$. Additional rates near the one yielding the highest feasible TSO benefit are then evaluated at progressively smaller intervals. This process continues until the prescribed resolution is reached, and the optimal sharing rate is found and its corresponding VSC configuration are returned.

Algorithm~\ref{alg:sharing_search} summarizes this procedure. Here, $\Delta\beta^{(0)}>0$ is the initial search step, $\gamma>1$ is the step-reduction factor, and $\varepsilon_\beta>0$ is the stopping threshold for the search step. In each round, $\operatorname{grid}(I,\Delta\beta)$ samples the current interval $I$ with spacing $\Delta\beta$. After each round, the step is divided by $\gamma$. The search stops once a round with $\Delta\beta\leq\varepsilon_\beta$ has been completed. Since sampling grids at different resolutions may share some candidate points, previously evaluated results are reused whenever these points are encountered again. If no feasible candidate has been found, the full interval $[0,1]$ is retained for the next round, provided the stopping threshold has not yet been reached.

\begin{algorithm}[!t]
\caption{Benefit-sharing-rate $\beta$ search for MTDC investment}
\label{alg:sharing_search}
\small
\DontPrintSemicolon
\KwIn{$\Delta\beta^{(0)}$, $\gamma$, $\varepsilon_\beta$}
\KwOut{Optimal benefit-sharing rate $\beta^\star$ and VSC configuration
$\bigl(\{z_{vi}^{\star}\},\,\{s_c^{\mathrm{vsc},v\star}\},\,\{n_c^{\mathrm{vsc},v\star}\}\bigr)$}
$I\gets[0,1]$, $\Delta\beta\gets\Delta\beta^{(0)}$, $\hat F\gets0$\;
\While{true}{
  \ForEach{$\beta\in\operatorname{grid}(I,\Delta\beta)$}{
    Solve the Inner Problem (MISOCP in Eq.~\eqref{eq:inner_planning} with fixed $\hat\beta$), or reuse its cached result\;
    \If{a feasible result improves the best saved solution}{
      Save $\hat\beta$, $\hat F$, and the VSC configuration $\bigl(\{\hat{z}_{vi}\},\,\{\hat{s}_c^{\mathrm{vsc},v}\},\,\{\hat{n}_c^{\mathrm{vsc},v}\}\bigr)$\;
    }
  }
  \lIf{$\Delta\beta\leq\varepsilon_\beta$}{\textbf{break}}
  \If{a feasible solution has been found}{
    $I\gets[\max\{0,\hat\beta-\Delta\beta\},
             \min\{1,\hat\beta+\Delta\beta\}]$\;
  }
  $\Delta\beta\gets\Delta\beta/\gamma$\;
}
\end{algorithm}

\section{Case Study and Discussion}
\label{sec:4}
We use the test system shown in Fig. 2 to validate our proposed planning model and solving method. The topology and line impedance of the 9-bus AC network are based on a simplified representation of the Slovenian 400kV transmission system, while the remaining parameters (e.g., base load levels, generator cost coefficients, transmission capacity, investment costs, etc.) are modified or reasonably assumed to suit the purpose of this study. A 3-terminal MTDC is considered for investment, with the AC bus connection and the capacity of each VSC determined by the proposed parametric model.

To facilitate analysis, annual system operation is simplified into four seasonal operating states. Each season comprises 2,190 hours, and the load is assumed to remain constant throughout the seasonal scenarios. The loads in spring, summer, autumn, and winter are set to 0.90, 1.05, 0.95, and 1.10 times the base load level, respectively. The generation limits, generator cost coefficients, and network parameters remain unchanged across seasonal scenarios. The VSC siting and sizing decisions are common to all seasonal scenarios, whereas the generation dispatch and power flow are optimized separately for each season. The key parameters involved in MTDC investment are listed in Table~\ref{tab:4}. The parametric MISOCP model and the associated sharing-rate search procedure formulated in Eq.~\eqref{eq:outer_incentiv} are implemented in Julia\footnote{\url{https://julialang.org/}}. JuMP\footnote{\url{https://jump.dev/JuMP.jl/stable/}} is used to formulate each fixed-$\beta$ MISOCP, which is subsequently
solved using Gurobi\footnote{\url{https://www.gurobi.com/product/}}.

\begin{figure}
    \centering
    \includegraphics[width=1\linewidth]{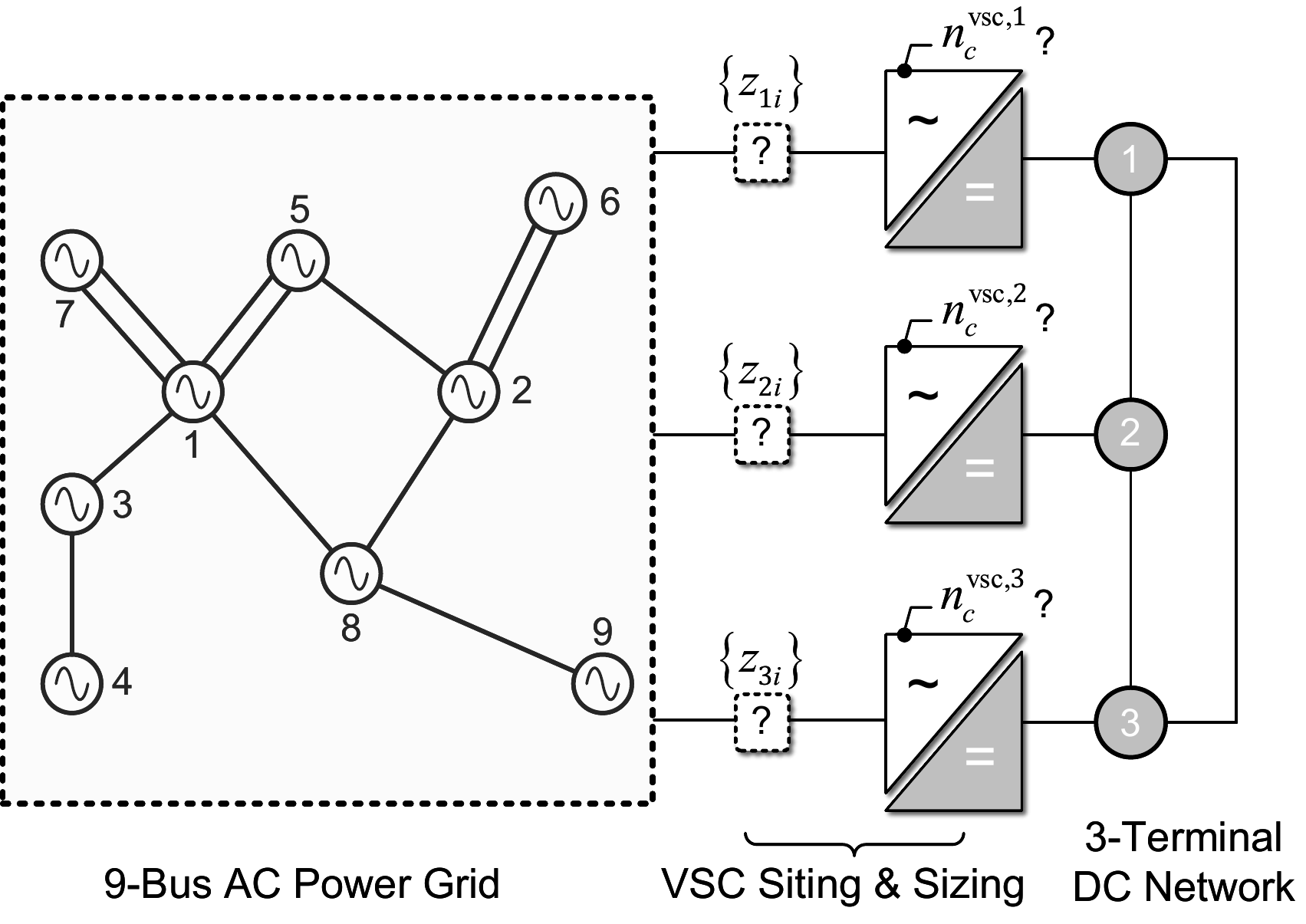}
    \caption{Customized AC/DC hybrid system. Each AC bus is considered a candidate connection point for the VSCs, whereas the MTDC network topology and the associations between the DC terminals and the VSCs are fixed. }
    \label{fig:2}
\end{figure}

\begin{table}[!t]
\centering
\caption{Assumed key parameters for MTDC investment}
\label{tab:4}
\renewcommand{\arraystretch}{1.10}
\setlength{\tabcolsep}{4pt}
\footnotesize
\begin{tabularx}{\columnwidth}{
    @{}X>{\raggedleft\arraybackslash}p{2.3cm}@{}
}
\toprule
Parameter and symbol & Value \\
\midrule

Annualized MTDC fixed cost, $K^{\mathrm{dc}}$
& $\$300{,}000$/yr \\

Annualized fixed cost per VSC, $K^{\mathrm{vsc},v}$
& $\$50{,}000$/yr \\

Annualized VSC unit investment cost,
$\kappa^{\mathrm{vsc},v}$
& $\$10{,}000$/MVA/yr \\

Minimum acceptable investor profit,
$\underline{\Pi}_{\mathrm{inv}}$
& $\$500{,}000$/yr \\

\bottomrule
\end{tabularx}
\end{table}

\subsection{Seeking Cost-Effective Incentive with Benefit-Sharing Rate $\beta$}
For each candidate benefit-sharing rate $\beta$, the corresponding fixed-$\beta$ MISOCP is solved to jointly determine VSC siting and sizing. The search starts with $\Delta\beta^{(0)}=0.1$ and uses a step reduction factor of $\gamma=5$. It terminates when the search step satisfies $\Delta\beta\leq\varepsilon_\beta$, where $\varepsilon_\beta=0.01$. This condition is met in the third round, giving search steps of $0.1$, $0.02$, and $0.004$, respectively.

\begin{figure}
    \centering
    \includegraphics[width=1\linewidth]{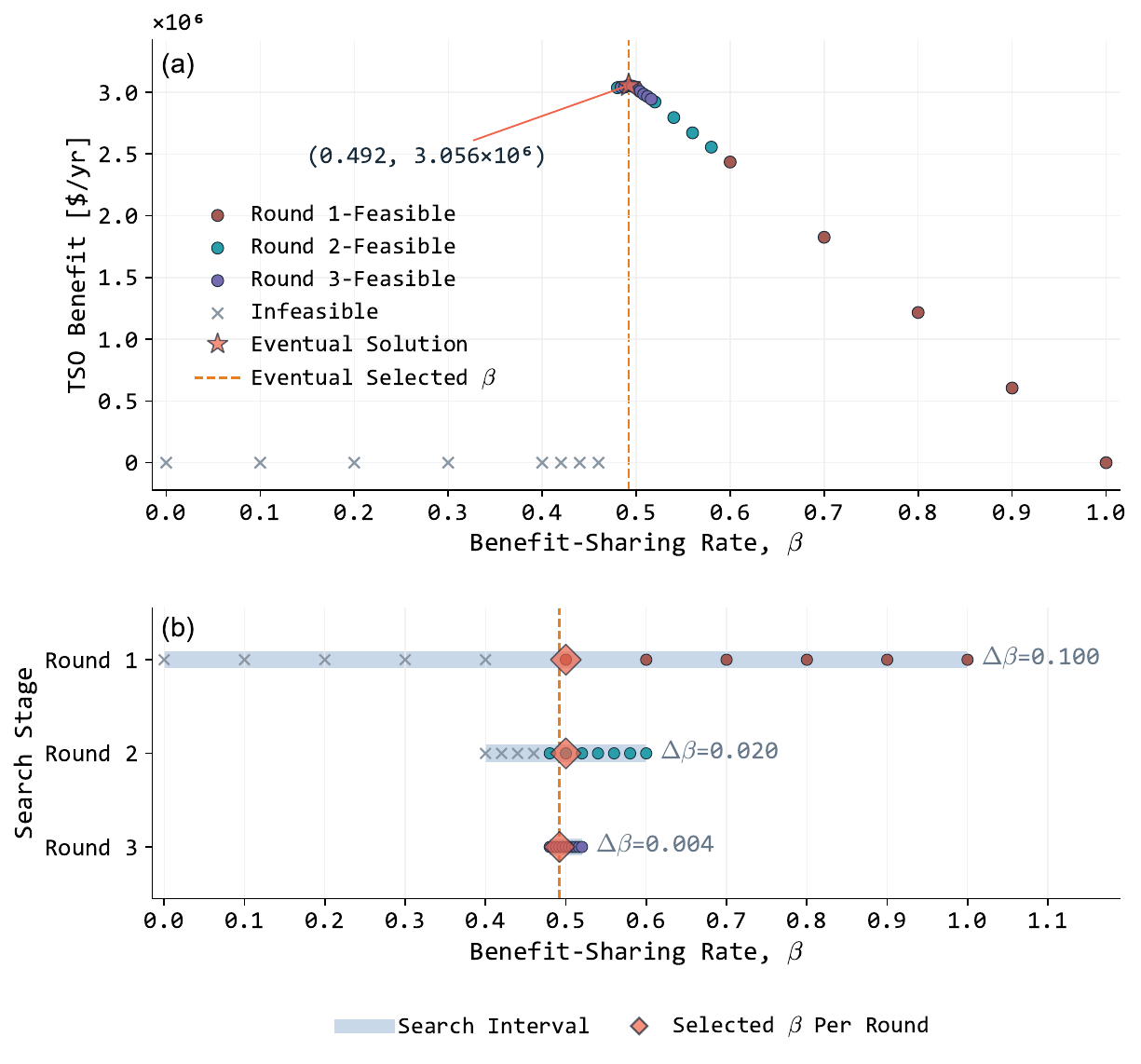}
    \caption{Search $\beta$ via Algorithm~\ref{alg:sharing_search}. (a) shows TSO benefits $F(\beta)$ across search rounds. (b) presents the candidate $\beta$ in each search round. }
    \label{fig:3}
\end{figure}

Figure~\ref{fig:3} and Table~\ref{tab:5} jointly illustrate the three-round search process. The selected sharing rate $\hat\beta$ remains at $0.500$ in the first two rounds and changes to $0.492$ after the final refinement, yielding a TSO benefit of $\$3.056\times10^{6}$/yr. During the multi-round search, some candidate values of $\beta$ reappear in subsequent rounds. Their previously obtained solutions are reused rather than recomputed, thereby avoiding unnecessary computational effort.

\begin{table}[!t]
\centering
\caption{Summary of $\beta$ search in three rounds}
\label{tab:5}
\footnotesize
\setlength{\tabcolsep}{3.5pt}
\renewcommand{\arraystretch}{1.12}
\begin{tabular}{@{}cccccc@{}}
\toprule
Round & Interval & Step & Candidates & New solves & Best $\beta$ \\
\midrule
1 & $[0.000,\,1.000]$ & $0.100$ & 11 & 11 & $0.500$ \\
2 & $[0.400,\,0.600]$ & $0.020$ & 11 &  8 & $0.500$ \\
3 & $[0.480,\,0.520]$ & $0.004$ & 11 &  8 & $0.492$ \\
\bottomrule
\end{tabular}
\end{table}

\subsection{Benefits of MTDC Integration}
At the selected $\hat\beta=0.492$, the resulting VSC siting and sizing results are presented in Table~\ref{tab:6}. Table~\ref{tab:7} further presents the benefits of integrating MTDC into the existing AC grid. Integrating the optimally configured MTDC network reduces annual generation costs by \$6,016,437, corresponding to a 2.85\% reduction. Although the percentage reduction is modest, the absolute annual savings are substantial.  At the selected $\hat\beta$, this benefit is distributed between the TSO and the MTDC investor. As a result, the investor's minimum-profit requirement is met while the TSO also obtains considerable economic benefits.
\begin{table}[!ht]
\centering
\caption{Selected VSC locations and installed capacities}
\label{tab:6}
\footnotesize
\setlength{\tabcolsep}{8pt}
\renewcommand{\arraystretch}{1.05}
\begin{tabular}{@{}ccc@{}}
\toprule
VSC & AC bus & Installed capacity [MVA] \\
\midrule
1 & 1 & $47$ \\
2 & 9 & $99$ \\
3 & 5 & $55$ \\
\bottomrule
\end{tabular}
\end{table}

\begin{table}[!t]
\centering
\caption{Comparison of annual generation costs }
\label{tab:7}
\footnotesize
\setlength{\tabcolsep}{3pt}
\renewcommand{\arraystretch}{1.15}
\begin{tabular}{@{}cccc@{}}
\toprule
\shortstack{Without MTDC\\{[\$/yr]}} &
\shortstack{With MTDC\\{[\$/yr]}} &
\shortstack{Savings\\{[\$/yr]}} &
\shortstack{Reduction\\{[\%]}} \\
\midrule
$211{,}145{,}160$ &
$205{,}128{,}723$ &
$6{,}016{,}437$ &
$2.85$ \\
\bottomrule
\end{tabular}
\end{table}

\subsection{Gap Between SOCP- and NLP-Based Optimal Power Flow Model}
After obtaining the optimal VSC siting and sizing decisions, we fix the siting and sizing of three VSCs. Based on this configuration, two snapshot-kind AC/DC OPF models are formulated under the base-load condition, along with the optimization objective set to minimize the generation costs.

\begin{itemize}
    \item SOCP-based OPF model: The AC/DC power flow and VSC operating constraints are represented by the SOCP relaxation used in the proposed planning model. 
    \item NLP-based OPF model \cite{b9}\footnote{The constraint formulations can be found in the ACDC-OpFlow Manual: \url{https://github.com/CRESYM/ACDC_OPF}.}: The corresponding original nonlinear AC/DC power flow and VSC operating constraints are retained.
\end{itemize}

As shown in Table~\ref{tab:8}, the generation-cost and total-generation gaps relative to the NLP results are minor. The close agreement supports the use of the SOCP relaxation for the system-wide power flow and VSC operating constraints in the proposed planning model.

\begin{table}[!t]
\centering
\caption{Comparison of SOCP- and NLP-based OPF results}
\label{tab:8}
\footnotesize
\setlength{\tabcolsep}{5pt}
\renewcommand{\arraystretch}{1.15}
\begin{tabular}{@{}lrrr@{}}
\toprule
Quantity & SOCP & NLP & Gap [\%] \\
\midrule
Generation cost [\$/h]
& $23{,}369$ & $23{,}384$ & $0.0619$ \\
Total generation [MW]
& $1315.40$ & $1316.06$ & $0.0496$ \\
\bottomrule
\end{tabular}
\end{table}

\section{Conclusion}
In this work, we investigate the expansion of an existing AC grid by integrating an MTDC network. The planning problem jointly determines the siting and sizing of the VSC stations. Crucially, it also addresses how to attract MTDC investment by sharing the resulting system benefits between the TSO and the MTDC investor. To this end, we develop a parametric MISOCP model and an associated search procedure for the cost-effective incentive related to benefit-sharing rate. Numerical experiments on a customized AC/DC case system demonstrate the effectiveness of the proposed model and solution method.

The present model provides only a preliminary representation of the interaction between the MTDC investor and the TSO. Future work will examine richer strategic and market behaviors, formulate them mathematically, and extend the optimization framework to better reflect practical investment decisions.

\label{sec:5}

\section{Acknowledgements}
This work was supported by the CRESYM project Harmony (\url{https://cresym.eu/harmony/}).

\end{document}